\documentclass[]{article}
\usepackage[preprint]{neurips_2022}
\usepackage{geometry}
\usepackage{enumitem} 
\usepackage{graphicx}
\usepackage[font=small,skip=0pt]{caption}
\usepackage{color}
\usepackage{amssymb}
\usepackage{amsmath}
\usepackage{amsfonts}
\usepackage{chronology}
\usepackage[utf8]{inputenc} 
\usepackage[T1]{fontenc}    
\usepackage[colorlinks=true,urlcolor=blue,linkcolor=blue, citecolor=black]{hyperref}
\usepackage{cleveref}
\usepackage{multicol}
\usepackage{url}            
\usepackage{booktabs}       
\usepackage{amsfonts}       
\usepackage{nicefrac}       
\usepackage{microtype}      
\usepackage{xcolor}         
\usepackage{ulem}
\usepackage{makecell}
\usepackage{paralist}

\expandafter\let\csname ver@natbib.sty\endcsname\relax

\usepackage[backend=biber,style=numeric,citestyle=numeric,sorting=none]{biblatex}
\newcommand{\comments}[1]{\textcolor{red}{\textbf{#1}}}

\title{ESA-Ariel Data Challenge NeurIPS 2022: \\
    \textbf{Inferring Physical Properties of Exoplanets From Next-Generation Telescopes} }
\author{Kai Hou Yip \thanks{kai.yip.13@ucl.ac.uk} \and I.P. Waldmann \and Q. Changeat \and M. Morvan \and A. Al-Refaie \and B. Edwards \and N. Nikolaou \and A. Tsiaras \and  C. Alves de Oliveira \and P.-O. Lagage \and C. Jenner \and J. Cho \and J. Thiyagalingam \and G. Tinetti \and
{\tt kai.hou.yip@ucl.ac.uk}\\
}
\date{\today}

\begin{document}
\maketitle

\begin{abstract}
The study of extra-solar planets, or simply, exoplanets,  planets outside our own Solar System, is fundamentally a grand quest to understand our place in the Universe. Discoveries in the last two decades have re-defined our understanding of planets, and helped us comprehend the uniqueness of our very own Earth. In recent years the focus has shifted from planet detection to planet characterisation, where key planetary properties are inferred from telescope observations using Monte Carlo-based methods. However, the efficiency of sampling-based methodologies is put under strain by the high-resolution observational data from next generation telescopes, such as the James Webb Space Telescope and the Ariel Space Mission. We are delighted to announce the acceptance of the Ariel ML Data Challenge 2022 as part of the NeurIPS competition track. The goal of this challenge is to identify a reliable and scalable method to perform planetary characterisation. Depending on the chosen track, participants are tasked to provide either quartile estimates or the approximate distribution  of key planetary properties. To this end, a synthetic spectroscopic dataset has been generated from the official simulators for the ESA Ariel Space Mission. The aims of the competition are three-fold. 1) To offer a challenging application for comparing and advancing conditional density estimation methods. 2) To provide a valuable contribution towards reliable and efficient analysis of spectroscopic data, enabling astronomers to build a better picture of planetary demographics, and 3) To promote the interaction between ML and exoplanetary science. The competition is open from 15th June and will run until early October, participants of all skill levels are more than welcomed!

\end{abstract}

\subsection*{Keywords}
Density Estimation, Probabilistic Inference,  Physical Modelling, Exoplanetary Atmosphere

\subsection*{Competition type} Regular

\section{Competition description}

\subsection{Background and impact}
\label{sec:background}
\begin{figure}[t]
    \vspace{-5pt}
    \centering
    \includegraphics[width=0.8\textwidth]{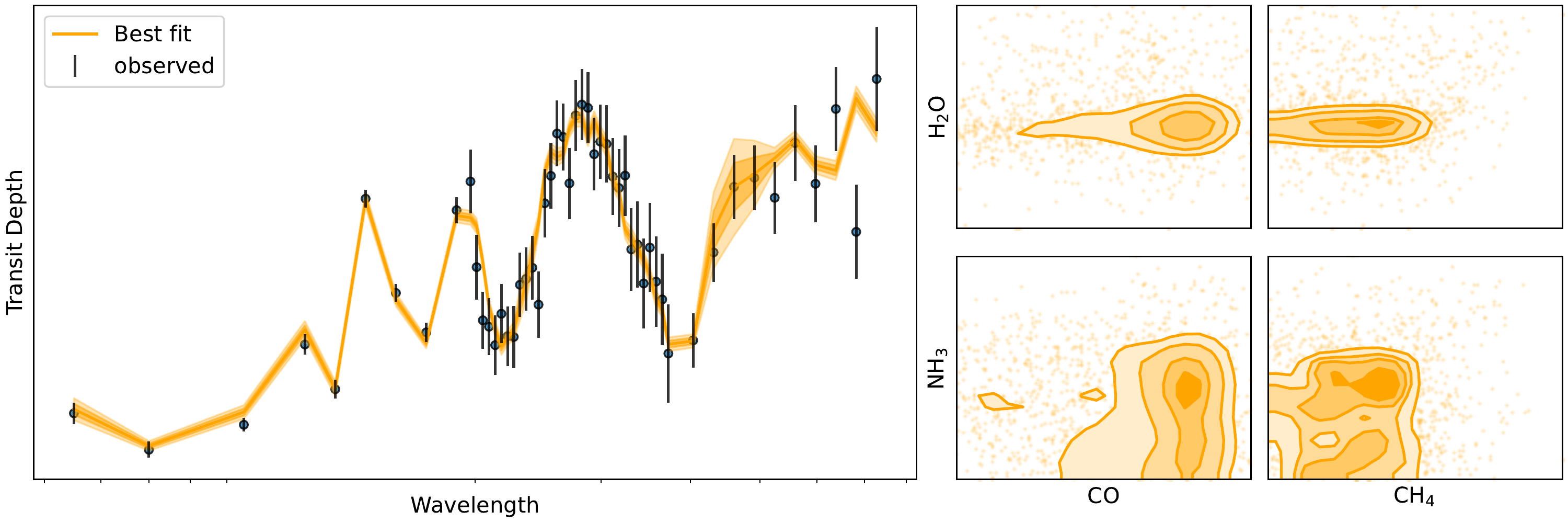}
    \caption{Example of the data provided. Left: An example exoplanet spectrum (black)  of the super-Earth  K2-18\,b, simulated for the Ariel Space Mission with the best fit atmospheric model superimposed (yellow). Right: Approximate conditional distribution of four atmospheric molecules. This is an example of the desired output (specifically, the interactions on the right) of this competition. }
    \label{fig:example_spectrum}
    \vspace{-15pt}
\end{figure}

In the last two and a half decades, we have undergone what is best described as a second Copernican revolution. The discovery of extrasolar planets \parencite{Mayor1995} - i.e. planets orbiting other stars - garnered the Nobel Prize in Physics 2019 \parencite{nobel_2019} and has fundamentally transformed our understanding of planets and our place in the Milky Way. Today, despite the many limitations of our detection techniques, we know of over 5,000 worlds outside our own solar system\footnote{ \url{https://exoplanetarchive.ipac.caltech.edu/}}. In other words, our Milky Way hosts 100s of billions of planets \cite{Cassan_2012}. Furthermore, and contrary to expectations, we now know that most of these planets are Earth-sized or so-called super-Earths (planets roughly between 1 - 10 Earth masses, \cite{Howard_2010, Fulton_2017, Zhu_2021}). These discoveries pose some fundamental questions: “How do planets form?”, “Are these worlds habitable?" and "Are we alone?” - To date, the answer is somewhat sobering: “We  really don't know”.

The most promising approach to answering the above questions, lies in studying their atmospheres. As the planets eclipse their host-stars (in our line of sight), a minuscule fraction of starlight (10 to 50 photons per million) passes through their atmospheric annuli and interacts with the chemistry, clouds and winds of the planet's atmosphere. Measuring and interpreting these faint signatures has recently been set as NASA's main scientific goal for this decade (see NASA Decadal Survey \parencite{NAP26141}). The recent launch of the James Webb Space Telescope (JWST), the upcoming Ariel Space Mission\footnote{\url{https://arielmission.space}}  \parencite{ariel}, and the Twinkle Space Telescope \parencite{twinkle}, all will collectively provide high-quality data on over 1,000 exoplanets in our galactic neighbourhood. 

Yet, there are a number of challenging problems to solve. Astronomers have relied on Bayesian posterior distributions to understand the impact of different atmospheric phenomena on the observed spectrum (see \Cref{fig:example_spectrum}, for an example spectrum and corresponding data analysis). Such analyses have led to many surprising results in the field of exoplanet atmospheres, such as the ubiquitous presence of water vapour \cite{Tinetti_water, Tsiaras_2018_pop, k2-18b_tsiaras}, the discovery of clouds \cite{Sing_2016, Anisman_2020}, or even some evidence of a connection between the composition of atmospheres and the way planets form \cite{Welbanks_2019, Changeat_2022} Increasingly complex planetary models and rising numbers of detected exoplanets will soon prohibit the use of Bayesian sampling approaches (i.e. MCMC, Nested Sampling, and alike.), creating a major bottleneck for data analysis \cite{Changeat_2021, alfnoor}. Lately, the field has begun to look at alternative methods, such as, ML-based approaches (e.g. \cite{cobb2019,random_forest_Neila,Fort_2019, lobato_2015,Izmailov_2021,Lakshminarayanan_2016}). These solutions often address overly simplified science cases, limiting their applicability to real-world scenarios. The field now needs, more than ever, a solution that can provide fast and reliable inference to datasets of substantial sizes. This competition is a first major step towards this effort. 


By examining the problem through the lens of exoplanet atmospheric characterisation, the inverse problem is infused with a number of challenging issues in which the NeurIPS community has significant expertise and experience, such as,

\begin{compactitem}
    \item Detecting signals in extremely low signal-to-noise ratio conditions,
    \item Identifying the presence of solution degeneracies (multiple equivalent solutions),
    \item Quantifyng input uncertainties,
    \item Dealing with limited or no ground truth cases (we cannot physically travel 100's of lightyears), and
    \item Handling data set shifts (due to instrument noise or seasonal variations).
\end{compactitem}

Inverse problems are ubiquitous in both industrial and academic settings.  Any outcome of this competition can thus be applicable to inference applications in any scientific/computational field. In particular, any solutions presented in this competition will feed directly into improving the analysis of Earth-analogue planets observed by JWST and Ariel.

\subsection{Previous Experience, Success and Future Outlook}

The team have had ample experience in running data challenges in the past and our platform, data generation and validation pipelines have been tried and tested. In 2019 and 2021 we ran two very successful data challenges as part of the ECML-PKDD "Discovery Challenges". We detail some of the key results of last-year's challenge below: 
\begin{compactitem}
    \item The largest ECML-PKDD challenge in the past three years (130 participants).
    \item The challenge received significant global media attention with the winner featured on Portugal's leading quiz show, Joker\footnote{\url{https://www.youtube.com/watch?v=4O1sLym-RD0}}.
    \item Three leading academic institutes used our data challenge as Masters thesis projects.
    \item The European Planetary Science Congress (the largest planetary science conference worldwide) now issues the Ariel Data Challenge Award to foster interdisciplinary collaborations between the Planetary Sciences and the AI/ML community.
\end{compactitem}

As is now customary, winners of this challenge will be invited to give talks at the Ariel Consortium Conferences hosted at the European Space Agency (ESA) or partnering institutions. We will also issue a joint ESA/UK Space Agency press release promoting the winners.

In the longer term, we plan to make the Ariel Data Challenge an annual fixture at either NeurIPS, ECML-PKDD or similar venues.

\subsection{Novelty}
The problem of inverse retrieval of exoplanet atmospheres has never been featured in any Data Science or ML-oriented data challenge or competition. Hence the application area for this competition is novel. Wilson \textit {et al.}'s NeurIPS competition in 2021\footnote{\url{https://izmailovpavel.github.io/neurips_bdl_competition/}} (W21) is perhaps the most similar conceptually to the proposed one, as they are both concerned with conditional probability estimation. Here we list our key differences: 

\begin{compactitem}
\item \textbf{Constraints}: W21 was intended as a comparison of sampling approaches for Bayesian Deep Learning methods on pre-defined model-dataset pairs. We impose no constraints on the method used to obtain the conditional density estimates.  
\item \textbf{Dataset}: W21 used pre-existing datasets consisting of 1 or 2 target variables, whose distribution given the inputs need to be estimated. The dataset used in this competition is specifically designed for multi-dimensional conditional density estimation and involves a 6-dimensional target space.
\item \textbf{Ground Truth}:
Ground truth predictive distributions in W21 were generated by running Hamiltonian Monte Carlo algorithms on pre-specified neural architectures. Our reference distributions are generated by an atmospheric physics model. Hence, our resulting distributions capture non-linear physically motivated relationships between targets. Any optimal submission must demonstrate the ability to learn physically meaningful relations/correlations. 
\end{compactitem}


\subsection{Data}
\subsubsection{Data Generation}
\label{sec:data_generation}
\begin{figure}[t]
    \vspace{-5pt}
    \centering
    \includegraphics[width=0.9\textwidth]{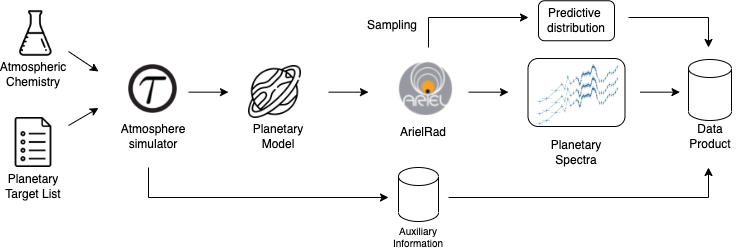}
    \caption{Schematic of Data Generation Procedure. }
    \label{fig:data_gen}
    \vspace{-15pt}
\end{figure}
This competition relied on simulated data generated from our dedicated physical forward model and instrument simulator. A total of 105,887 forward models has been produced and 26,109 of them will be complementary retrievals results coming from Nested Sampling. Data generation of this scale has never been attempted before due to computational intensive nature of retrievals . We circumvented the constraints with our fully automated and parallelizable data generation pipeline \cite{alfnoor}. The exclusive use of simulated data stemmed from the fact that only hundreds of planets have been characterised with spectroscopy, far below the minimum threshold for any training meaningful machine learning models. Below we will brief describe the process and key considerations behind each step. Interested reader could refer to \cite{ABC_database} for a much more in-depth discussion on the data generation aspect of this competition. The corresponding database, the Ariel Big Challenge (ABC) Database, is made permanently available at \url{https://doi.org/10.5281/zenodo.6770103}. 
Below, we briefly describe the steps involved in this process (Also see \Cref{fig:data_gen} for a summary). 

\noindent\textbf{Step 1: Planet Selection} \newline
We selected observable planets from currently confirmed exoplanet and TESS candidates, leaving us with about 6000 objects\footnote{We decided not to supplement the list with synthetic planets as it remains unclear how different classes of planets are formed in the first place.}. They were divided randomly into a training and test set. Planets of either set are subjected to the same treatment.

\noindent\textbf{Step 2: Atmospheric Synthesis} \newline
We simulated different instances of planetary atmospheres for each planet. Their atmospheres were injected with 5 atmospheric gases (H$_2$O, CH$_4$, CO, NH$_3$, CO$_2$) at abundances that are randomized, yet consistent with leading planetary formation theories \cite{Moses_2012,madhusudhan_atmospheric_2017}.

\noindent\textbf{Step 3: Transformation into Observations} \newline
We convolved our simulated spectra with the instrument response from the ESA Ariel Mission radiometric model, ArielRad \parencite{mugnai_Arielrad}.

\noindent\textbf{Step 4: Ground Truth Generation } \newline
We used the MultiNest algorithm \cite{feroz2008, py_multinest} to generate a library of approximate 6-dimensional conditional distributions (5 atmospheric gases \& planet temperature) as our ground truths\footnote{We note that these distributions are \emph{not} the actual ground truth, but they currently represent our best estimates of the ground truth given the observational noise. }, using a Gaussian likelihood and a uniform prior. 

After data cleaning and some pre-processing steps, the final dataset includes 91,392 instances of simulated observations from the upcoming Ariel Space Mission, 21,988 of them ($\sim 24 \%$) are labelled (ground truth generated from step 4), while the rest are unlabelled to allow for both supervised and semi-supervised learning based solutions. Simulation of this scale has never been attempted before. It is achieved using the open-source software, \texttt{TauREx3} \parencite{taurex}  \parencite{TauREx31} and \texttt{Alfnoor} \parencite{alfnoor}. Both software frameworks were developed by members of this proposal and have been demonstrated in numerous peer reviewed publications.  

The final dataset is available to the community as part of the outcomes of this competition \footnote{\url{}}. The dataset, with its realistic observation uncertainties and well-curated list of known planets, can serve as an excellent benchmark dataset for a diverse array of research directions, be they domain specific, or for approximate interference method development in general.
\vspace{-2mm}
\subsection{Data Overview}
\begin{figure}[t]
    \vspace{-10pt}
    \centering
    \includegraphics[width=0.8\textwidth]{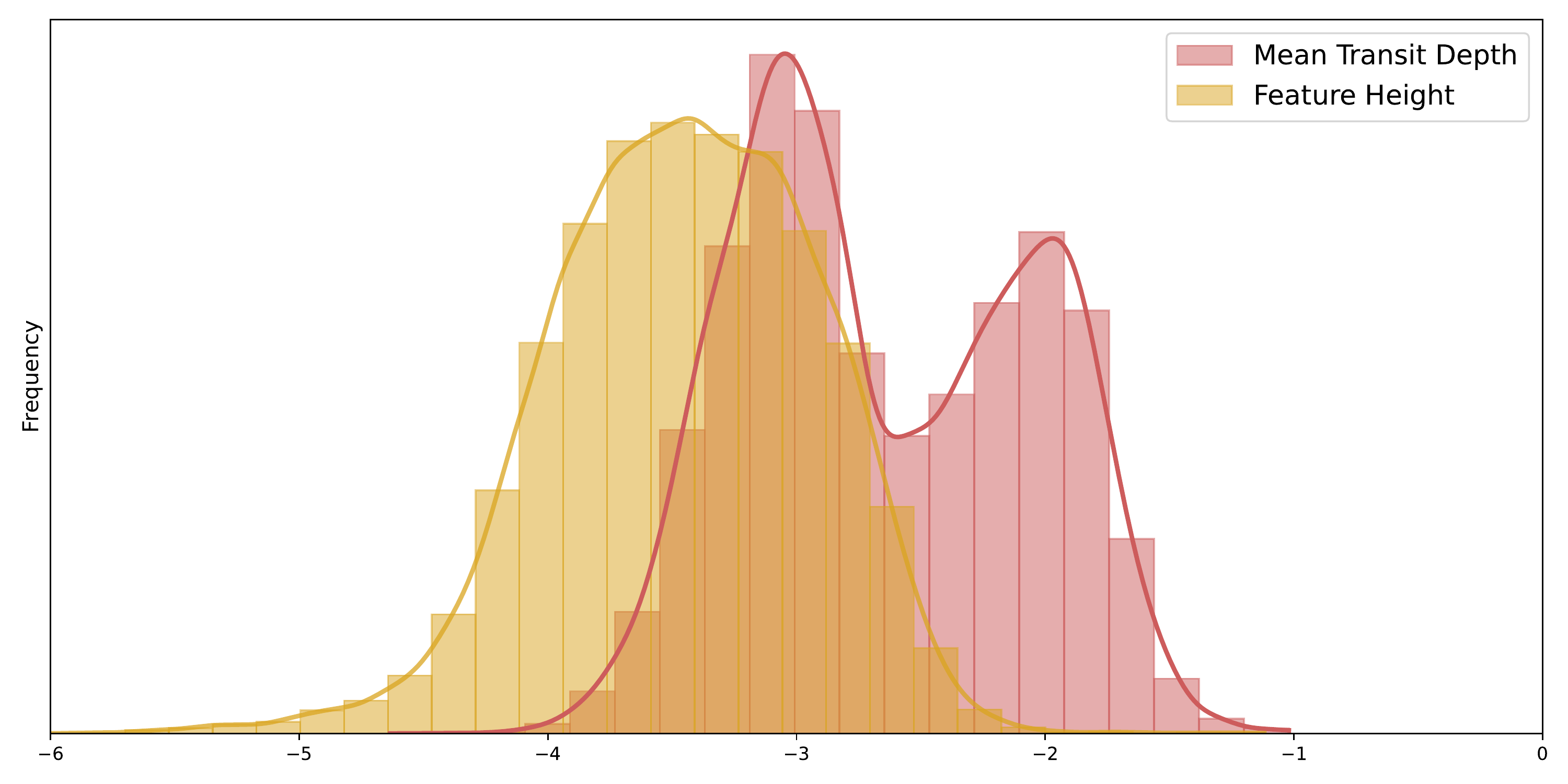}
    \caption{Distributions of aggregated planetary characteristics in the data set: Red: Distribution of mean transit depth,  Orange: Distribution of feature height. The $\sim$3 orders of magnitude range demonstrates a vast feature space with varying feature strength.   }
    \label{fig:data_distribution}
    \vspace{-15pt}
\end{figure}

Figure \ref{fig:data_distribution} provides an overview of the data product. Subplot (a) shows the distribution of (log-)mean transit depth, a proxy to measure the ``type'' of planet which can range from Earth-like to inflated ultra-hot Jupiters. This distribution illustrates a diverse feature space for this inference task. Subplot (b) shows the distribution of (log-)feature height, a first-order estimate on the ``strength" of the spectroscopic features. The $\sim$3 orders of magnitude range demonstrate the difficulty of having to build inference frameworks that are able to cope with large feature variations as well as inferring features close to the noise limit.


\subsubsection{Training Data}

The participants will receive three sets of data:
\begin{compactitem}
    \item \textbf{AuxillaryTable.csv}: a $N_{training} \times 9$ table. This includes all known and relevant astrophysical and instrumental information on an observation, such as star distance, stellar masses, stellar radii, stellar temperature, orbital periods, semi major axis, planet masses, planet radii and surface gravity of the planet.
    \item \textbf{SpectralData.hdf5}: a $N_{training} \times 4 \times N_{\lambda}$ table. Each training example contains spectroscopic information about the observation (wavelength , feature-intensities, observation error and spectral-bin resolution) across all 52 ($N_{\lambda}$) wavelength channels
    \item \textbf{Ground Truth Package}: The package will include three tables. (1) \textbf{Tracedata.hdf5}: Traces of Nested Sampling generated approximate conditional distributions. (2) \textbf{QuartilesTable.csv}: 16th, 50th and 84th percentile estimates for each target. (3) \textbf{FM\_Parameter\_Table.csv}: Input values of each target used during data generation phase.
\end{compactitem}
Note that (1) \textbf{Tracedata.hdf5} and (2) \textbf{QuartilesTable.csv} are provided for labelled points only ($\sim$24 \% of the training data set). Unlike (1) \& (2), \textbf{FM\_Parameter\_Table.csv} (3) is provided for the full training set (both labelled \& unlabelled datapoints, 91,392 of them) and can be useful for semi-supervised learning approaches. 

Each row in the table represents a single planet instance. $N_{training}$ and $N_{\lambda}$ refer to the number of training datapoints and the number of wavelength channels respectively. The participants are not constrained w.r.t. which parts of the training data are used to train their models and how.

\subsubsection{Test Data}
\label{sec:test_data}
As for the test data, participants will receive the \textbf{AuxillaryTable.csv} and the \textbf{SpectralData.hdf5} files, but not the Ground Truth Package.


Submissions is evaluated based on their ability to generalise within the training set distribution \textbf{In-Training Parameter Ranges} (In-Range) and the ability to extrapolate to  \textbf{Out-of-Training Range Parameters} (Out-Range). In-Range samples are drawn from a reserved set of the original training data.
Out-Range tests will test the algorithm's ability to extrapolate to new, unseen scenarios and the test data will be generated by simulating unseen planetary and atmospheric properties. Out-range parameters will be within $\lesssim 10 \%$ extrapolation of In-Range bounds.  The test set will consist of 4 (equal in size) subsets and the model's leaderboard score will be computed based on the average performance across these:
\begin{table}[h]
\centering
\begin{tabular}{lllll} 
\toprule
                        & Set 1           & Set 2               & Set 3               & Set 4               \\ \hline
Planetary Configuration & In-Range & Out-Range    & In-Range & Out-Range \\
Atmospheric Properties   & In-Range & In-Range & Out-Range     & Out-Range \\
\bottomrule
\end{tabular}
\vspace{-10pt}
\end{table}

In all cases, the ground truth distribution is generated with the default atmospheric model. There are in total 2997 test cases available, which means $\sim$750 examples for each set. Scores displayed in the public leaderboard will be computed based on 500 randomly selected test examples, and the final score will be computed using the remaining 2500 test examples. 

\subsection{Tasks and application scenarios}
\label{sec:tasks}
The main objective of the competition is to perform a supervised/semi-supervised multi-target inference given inputs from Ariel observation (SpectralData.hdf5) and auxiliary information (AuxillaryTable.csv). There are 6 targets in total: $T_p, X_{H_2O},X_{CH_4},X_{CO},X_{NH_3},X_{CO_2}$, where $T_p$ is the planet temperature and $X$ is the abundance of a given atmospheric gas. Below we will specify our tasks for the two tracks, the \textbf{Light Track} and the \textbf{Regular Track}. For both tracks the targets remain the same.


\textbf{Light Track:} Participants are asked to provide the 16th, 50th and 84th percentile estimates for each target and test case. Solutions could provide reliable, first-order error estimations on the physical parameters. This helps speed up the model selection process for field practitioners. 

\textbf{Regular Track:} Provide an approximate 6-dimensional conditional distribution for all test cases. Solutions could serve as a direct replacement to the existing pipeline, allowing fast and efficient inference for large samples of planets. 


\subsection{Metric}
For both tracks, each dimension $t$ in our target multivariate distribution is normalised according to their respective prior bound.


In terms of the \textbf{Regular Track}, we choose to evaluate the similarities using the Wasserstein-2 distance, $W_2$. The Wasserstein-2 distance, also known as Earth-Mover distance, is a metric stemmed from the theory of optimal transport and is used to evaluate the level of overlap between two multivariate distributions, with 0 indicating maximally similar (i.e. the same) distributions and 1 maximally dissimilar.
\begin{equation}
     W_{2,n}(F_n, \hat{F}_n) = \underset{\pi\in \Gamma(F_n, \hat{F}_n)}{\inf} \int_{\mathbb{R}\times\mathbb{R}} |x-y|d\pi(x,y)
\end{equation}
where $F_n$ and $\hat{F}_n$ represent the Nested Sampling (NS) generated approximate conditional distribution and the participant's surrogate distribution for a single test case (n) respectively. $\Gamma(F_n, \hat{F}_n)$ represents the set of probability distributions on the metric space $\mathbb{R}\times\mathbb{R}$, whose marginal distributions are $F$ and $\hat{F}$ on first and second factor respectively.
The overall score of each participant will be the average score they obtain over the entire test set,
\begin{equation}
   \bar{W}_2 = \frac{\sum_{n=1}^N W_{2,n}}{N}
    \label{eqn:w2_score}
\end{equation}
We subtract $\bar{W}_2$ by unity and multiply the result by 1000 to turn it into a monotonically increasing function
\begin{equation}
    \text{score}_{Regular} = 1000(1 - \bar{W}_2)
    \label{eqn:regular_score}
\end{equation}

For the \textbf{Light Track}, submissions will be evaluated by comparing distance between the quartiles estimates of each target $t$ given by Nested Sampling ($q_{l,t}$) and the submission ($\hat{q}_{l,t}$), where $q \in [1,2,3]$. For each test case we will compute the average relative RMSE out of all quartiles and targets, i.e.

\begin{equation}
    RMSE_{n} = \sqrt{\mathbb{E} \left(\frac{q_{l,t} - \hat{q}_{l,t}}{q_{l,t}}\right)^2}
 \label{eqn:low_q}
\end{equation}

The participants' performance on the Light Track will be measured based on the average performance over the entire test set, i.e. $\bar{S} = \frac{\sum^N_n RMSE_n }{N} $. We have added a factor of 1000 to amplify the changes in $\bar{S}$ and subtract the result from a constant (1000) to create a monotonically increasing score function.
\begin{equation}
    \text{score}_{Light} = 1000(1 - \bar{S})
    \label{eqn:light_score}
\end{equation}
Thus the higher the score, the better the performance. Winners will be selected based on the overall score of the final evaluation test set. 

\subsection{Baselines, code, and material provided}
\begin{table}[h]
\vspace{-5pt}
\centering
\begin{tabular}{@{}lll@{}}
\toprule
                       & Light Track          & Regular Track                      \\ \hline
MVN Gaussian               & \comments{611 / 1000   }               & 163 / 1000\\
MC Dropout               & 635 / 1000                  & 139 / 1000\\
\bottomrule
\vspace{-15pt}
\end{tabular}

\end{table}
Based on current state of the art in the field (\cite{cobb2019};\cite{Francisco_2022}), We provide two baselines, a Deep Multivariate Gaussian Model and a Monte Carlo Dropout Model, both capable of providing a 6-D distribution for each input spectrum. A detailed description of the network and the pre-processing steps will be made available to the participants as part of the starter kit released in June during the beta test period. The main data generation tool, \texttt{TauREx3}, is publicly available on Github\footnote{Documentation: \url{https://taurex3-public.readthedocs.io/en/latest/}}. We will also have a limited number of slots of HPC resources sponsored by DiRAC HPC facility (see resource section below).  

\subsection{Website, Tutorial \& documentation}
The competition will be hosted on \url{https://www.ariel-datachallenge.space/}. The format is similar to previous challenges\footnote{see \url{https://www.ariel-datachallenge.space/2021/} for our challenge in 2021}. We aim to make the website and its resources easily accessible to the participants. The website will be organised into the following areas: (1) Home (Front Page) (2) Getting Started (3) Leaderboard (4) Tutorials \& Documentation (5) Contact Us (6) FAQ. The backend is complete, tried and tested from the 2021 challenge and the new frontend design will be finished by May 1$^{st}$. Conceptually it will be very similar to the 2021 website (see footnote).

The data challenge platform will be run on a dedicated webserver (128 cores, 150 Tb space). Our hardware setup allows for real-time calculations of participant performance metrics and leaderboard updates even in the event of large number of simultaneous submissions.

To equip participants with sufficient background knowledge. We will design a jupyter-notebook tutorial to illustrate the data generation process, discuss any assumptions and demonstrate the atmospheric model. We will also provide a blog-style explanation of the problem and physics and a list of more in-depth background literature for participants. 




\section{Organizational aspects}
\subsection{Protocol}
In order to enter the competition, participants will need to create an account for themselves/their team. The data is publicly available from the website. To evaluate their results, the participants will only be submitting results (i.e. their models’ approximate conditional distribution/error estimates on the test set examples) to our online website. Entry for both tracks will be clearly marked to avoid confusion. These will be a collection of conditional distribution/error estimations as described in \Cref{sec:test_data}. We will automatically evaluate these predictions according to \Cref{eqn:regular_score} for the \textbf{Regular Track} and \Cref{eqn:light_score} for the \textbf{Light Track}, and report the submissions’ score on a leaderboard we will be maintaining on our webpage. The competition will not involve multiple phases and participants are free to make as many submissions as they wish to either or both tracks. The only restriction is that consecutive submissions cannot be made within 24 hours of one another. This is a measure to (i) prevent the participants from producing a trivially/randomly different model just to climb a bit higher on the leaderboard and (ii) to limit the extend to which the provided solutions overfit to the test set. These are both dangers due to the presence of the leaderboard which is effectively causing some information leakage from the test set. To further mitigate this danger, the leaderboard score will only be calculated on 50\% of the available test data (chosen uniformly at random from the full test set), while the final score for each solution (which will determine the final ranking of solutions) will be calculated on the full test set.

To prevent cheating, shortly after the end of the competition, the authors of the two top-ranked entries will be requested to provide a brief description of their solution (data preprocessing steps, model and training details) to the organizers (1–2 pages) and a copy of their code\footnote{We will offer participants the option of signing a Non-Disclosure Agreement before submitting their code and algorithmic description.}. Provided they do so and we do not identify any of the following issues:
\begin{compactitem}
\item The authors had access to ground truth.
\item The model relies heavily (as judged by the organizers) on hard-coded elements that are solely deemed to be due to test-set leakage.
\item The authors are participating in the competition under multiple aliases.
\end{compactitem}
Then these two participants will be deemed the two winners of the competition and receive their prizes. If any of the above issues are identified, the corresponding entrant will be disqualified and removed from the leaderboard, in which case the same procedure will be repeated on the remaining entries. 
The online platform will be operational by May 1$^{st}$ 2022. From this time until  May 30$^{th}$, members of the organising team and selected postgraduate students will beta test the platform.

\subsection{Rules}

\textbf{Draft of contest rules}
\begin{compactenum}
    \item Each participant must have a unique alias.
    \item No participant can submit more than 1 entry in a 24 hour interval. If they do, their second entry will be automatically ignored.
    \item After the end of the competition and before announcing the winners, the authors are required to provide a brief description of their solution to the organisers (1–2 pages) and a copy of their code. Failure to do so in the agreed upon time frame will lead to their disqualification.
    \item The organisers will use the provided code, models and solutions only for the purposes of checking whether the contest rules (5) \& (6) are not broken. The organisers will not use any result without the authors’ explicit permission.
    \item Participants must not have access to the ground truth before the competition’s closing date. If they do, they will be disqualified.
    \item For an entry to count as a winning entry, it must not rely heavily (as judged by the organisers) on hard-coded elements that are solely deemed to be due to test-set leakage.
    \item Participants are allowed to form teams. In this case, they must notify the organisers to deactivate their individual aliases and they must create a joint (group) alias. The group alias shall then be treated as a single participant for all purposes, including the restriction on the number of daily submissions and the prize received. All members of a winning team will be listed as winners of the competition and invited for further collaboration (e.g. for publishing the results with the organisers). However, a single prize will be awarded to the entire team. The team can -of course- decide how to receive the prize (e.g. nominate a single member to receive it, or split a monetary equivalent among them).

\end{compactenum}

The purpose of Rules (1) \& (2) is to prevent the participants from generating multiple trivial random variations of their models which would lead to overfitting to the test set score signal. To enforce Rule (1) we will request each participant to provide their full name, email address and affiliation (if applicable). We plan for Rule (2) to be automatically implemented in our online platform. They also allow for limiting the amount of traffic on the competition’s platform.

The aim of Rule (3) is to check whether the contest Rules (5) \& (6) are not broken. Rule (4) guarantees that the organisers will not use any result without the authors’ permission. If a submitted solution is novel and useful to the field, we plan to collaborate with its author (see \cite{Nikolaou_2019} as an example). Rules (5) \& (6) are measures for preventing cheating and overfitting to the test data. Rule (7) allows for and handles team formation.

To allow for the broadest possible participation, the set of rules is the minimal possible. There is no restriction on the models, algorithms or data pre-processing techniques, neither on the programming languages, environments or tools used for their implementation. The participants are also free to use data augmentation techniques, pre-trained models or any prior domain knowledge not included in the provided dataset. Finally, they are free to choose their own way of splitting the training data between training and validation sets.

\subsection{Competition promotion}

Our team of organisers comes from a diverse academic and cultural background, spanning across institutes in the UK, Europe and the USA. Our promotion campaign will leverage these networks to allow the broadest reach to our potential audience across the globe. A call for participation will be circulated in the following channels:

\begin{table}[h]
\resizebox{\columnwidth}{!}{%
\begin{tabular}{ll}
Astrophysics                        & Machine Learning           \\ \hline
European Space Agency (ESA)         &    The DiRAC HPC Facility                                            \\
The Ariel Space Mission Consortium  &  The Alan Turing Institute                                         \\
United Kingdom Space Agency (UKSA)  & SciML Research Group, RAL STFC                  \\
CCA, Simons Foundation &     Connectionists mailing list   \\
Instituto Nazionale di Astrofisica &  UCL Centre for Data Intensive Science and Industries\\
Centre à l'Energie Atomique (CEA) & UCL Computer Science  \\
UCL Physics and Astronomy &    Machine Learning News group     \\
UK Exoplanet mailing list  &  ALLSTATS mailing list  \\
NASA Exoplanet mailing list & Machine learning for Planetary Science (ML4PS) \\
JWST exoplanet mailing list &   Royal Statistical Society    \\
European Planetary Science Congress mailing list&  \\
Europlanet Society magazine and mailing list & \\

\end{tabular}%
}
\vspace{-10pt}
\end{table}

\subsection{Resources provided by organizers, including prizes}
\vspace{-5pt}
The 1$^{st}$ and 2$^{nd}$ place winners from both tracks will be awarded \$2,000 and \$500 for the first and second prizes respectively
The organisers have access to sufficient resources for generating the dataset and evaluating solutions.

DiRAC\footnote{\url{https://dirac.ac.uk}} will sponsor access to GPU computing resources for $~$20 participants. These sponsored computing resources are aimed at participants from `underpriviledged' backgrounds without access to sufficient HPC resources to participate otherwise. These sponsored places will be made available through a short application questionnaire on our data challenge website and will be offered on a first-come-first-served and case-by-case basis.

\subsection{Support requested}
\vspace{-5pt}
We do not require any support from the conference.

\section{Resources}

\subsection{Organizing team}

Dr. Kai Hou (Gordon) Yip, \textit{University College London} (coordinator,evaluator,beta-tester), is a Research Fellow at UCL ExoAI group focusing on designing ML \& Explainable AI solutions for the modelling of exoplanet atmospheres. He is a coordinator of the Machine Learning working group within the ESA Ariel Space Mission consortium and an award holder for the Postdoctoral Enrichment Award from the Alan Turing Institute. During his PhD he has employed statistical methods to characterise planetary atmospheres. He also works on generative adversarial models for detecting and simulating ultra-low signal-to-noise observations of extrasolar planets. 

Dr. Ingo Waldmann, \textit{University College London / The Alan Turing Institute} (coordinator, evaluator), is an Associate Professor in astrophysics at UCL, deputy director of the Centre for Space Exochemistry Data (CSED), a Turing Fellow at the Alan Turing Institute and the principal investigator (PI) of the European Research Council project ExoAI. The ExoAI group focuses on designing and implementing machine learning approaches in the data analysis and modelling of exoplanet and solar-system data. He is leading the Machine Learning work of the  European Space Agency Ariel mission consortium. He is furthermore the UCL representaive on AI to the European Space Agency (ESA-labs) and UCL AI steering board member. He is a director and co-founder of Spaceflux Ltd focusing on AI applications in global robotic telescope space situational surveillance. 

Dr. Quentin Changeat, \textit{University College London} (data provider), is a post-doctoral research fellow at UCL and an expert in modelling and data interpretation of exoplanet atmospheres. He is a leader in the analysis of data from the Hubble Space Telescope and has exploited this instrument to understand the chemistry and thermal structure of many exoplanets. Dr. Quentin Changeat is also the coordinator of the Spectral Retrieval working group for the future ESA telescope Ariel. In September 2022, he will join the Space Telescope Institute as an ESA Fellow to work on the analysis of exo-atmospheres with the NASA James Webb Space Telescope. The codes he developed will be used to generate the datasets of this data-challenge.

Dr. Billy Edwards, \textit{Commissariat à l’Energie Atomique et aux Energies Alternatives} (data provider), is a post-doctoral fellow at CEA. His work has a general focus on exoplanets but with particular emphasis on spacecraft capability and operations studies. He works on modelling the performance of the JWST, Ariel and Twinkle (\url{http://www.twinkle-spacemission.co.uk}) space missions, exploring their expected capabilities and optimising target selection. We will base our simulated planets on his Ariel target list. 

Mr. Mario Morvan, \textit{University College London} (beta tester, evaluator), is a PhD candidate at the UCL Centre for Doctoral training in Data Intensive Science. His focus is on developing innovative deep learning techniques to denoise astronomical time series and characterise exoplanet transits from various space-based missions. Mr. Morvan provided the baseline model for our 2019 and 2021 ECML-PKDD Data Challenge. 
Dr. Ahmed Al-Rafaie, \textit{University College London} (website, server-backend, platform administrator), is the head of numerical simulation at UCL-CSED. He will lead the web-portal design and implementation as well as contribute to computational aspects of the data set simulations. 

Dr. Angelos Tsiaras, \textit{Instituto Nazionale di Astrofisica} (data-provider), is a senior research fellow at UCL. He is a leading expert in data analysis for Hubble Space Telescope data. He is the lead author of the first detection of a potentially habitable atmosphere around a temperate super-Earth \parencite{k2-18b_tsiaras} and he authored the largest self-consistent catalogue of exoplanet atmospheres to-date. Dr. Tsiaras is leading the data generation effort. He is one of the co-organiser of our 2019 and 2021 ECML-PKDD Data Challenge. 

Dr. Nikolaos Nikolaou, \textit{University College London} (consultant, evaluator, beta-tester), is an Assistant Professor in Data Intensive Science. His background is in machine learning and especially feature selection, ensemble methods \& cost-sensitive learning. His current interests involve causal inference, explainable machine learning methods for astrophysics and pharmaceutical \& medical applications. He was the main organiser of our 2019 ECML-PKDD Data Challenge and a co-organiser of our 2021 ECML-PKDD Data Challenge. \\

Dr. Catarina Alves de Oliveira, \textit{European Space Agency} (consultant), Catarina is a Science Operations Development Manager, she is responsible for the definition, design and implementation of the ESA contribution to science operations of the Ariel and SMILE missions. She is responsible for quality control of the simulation product from ArielRad.

Dr.~Pierre-Olivier Lagage, \textit{Commissariat à l’Energie Atomique et aux Energies Alternatives} (Instrument model provider), is a research director at CEA. He is co-Principal Investigator (co-PI) of the MIRI instrument on board of the JWST and is leading the MIRI European consortium program dedicated to JWST observations of exoplanet atmospheres. He is also co-PI of the Ariel mission and leads the Ariel science group in charge of the synergies between JWST and Ariel. He is the director of the CEA’s astrophysics department (staff around 200 people). 

Prof. James Cho, \textit{Center for Computational Astrophysics Flatiron Institute, Simons Foundation}, is a Research Scientist at CCA and Professor at Queen Mary University in London. His work focuses on modelling exoplanet atmospheric dynamics. In particular, he pioneered in the development of numerical methods to simulate complex exoplanet climates.

Dr. Clare Jenner, \textit{Distributed Research utilising Advanced Computing (DiRAC)} (consultant on HPC facility), is Deputy Director of the STFC DiRAC High Performance Computing Facility. She directs the facility’s training programmes supporting the prestigious DiRAC user-community’s educational, research and innovation activities. She will be providing advice on HPC resource management.

Dr. Jeyan Thiyagalingam, \textit{Rutherford Appleton Laboratory}, is the Head of Scientific Machine Learning (SciML) Research Group at Rutherford Appleton Laboratory, Science and Technology Facilities Council (UK),  and a Turing Fellow at the Alan Turing Institute UK (consultant on ML methodologies). The SciML group is one of the UK's premier AI for Science research groups, specialising on developing and applying AI/ML techniques for a wide range of problems across a range of scientific domains, including astronomy. He is also one of the general chairs for the MLCommons Science Working Group, an international body that leads AI Benchmarking initiatives to shape the AI landscape, especially with the focus on science. He also leads the AI Benchmarking for Science initiative across various US-UK national labs.

Prof. Giovanna Tinetti, \textit{University College London} (consultant), is a professor in astrophysics at UCL and the Principal Investigator (PI) and science lead of the Ariel space mission and scientific lead of the Twinkle Space Mission. Additionally, she is the PI of the ExoLights group at UCL working on Bayesian inverse retrievals of exoplanet atmospheric data and atmospheric modelling. Prof. Tinetti is the director of UCL’s Centre for Space Exoplanet Data (UCL-CSED). \\

\subsection{Resources provided by organizers, including prizes}

The 1st \& 2nd place winners from both tracks will be awarded \$2,000 and \$500 for the first and second prizes respectively
The organisers have access to sufficient resources for generating the dataset and evaluating solutions. \\

DiRAC\footnote{\url{https://dirac.ac.uk}} will sponsor access to GPU computing resources for $~$20 participants. These sponsored computing resources are aimed at participants from `underpriviledged' backgrounds without access to sufficient HPC resources to participate otherwise. These sponsored places will be made available through a short application questionnaire on our data challenge website and will be offered on a first-come-first-served and case-by-case basis.

\printbibliography
\end{document}